\documentclass[aps,pre,preprint,showpacs,showkeys,superscriptaddress,amsmath,amssymb]{revtex4}
\expandafter\let\csname equation*\endcsname\relax
\expandafter\let\csname endequation*\endcsname\relax
\usepackage{cleveref}
\usepackage{graphicx}
\begin{document}
\title[Transfer entropy between communities in complex networks]{Transfer entropy between communities in complex networks}
\author{Jan Korbel}
\email{korbeja2@fjfi.cvut.cz}
\affiliation{Department of Physics, Zhejiang University, Hangzhou 310027, P.~R.~China}
\affiliation{Faculty of Nuclear Sciences and Physical Engineering, Czech Technical University in Prague, B\v{r}ehov\'{a} 7, 115 19, Prague, Czech Republic}

\author{Xiongfei Jiang}
\affiliation{Department of Physics, Zhejiang University, Hangzhou 310027, P.~R.~China}
\affiliation{School of Information Engineering, Ningbo Dahongying University, Ningbo 315175, P.~R.~China}
\affiliation{Research Center for Finance Computing, Ningbo Dahongying University, Ningbo 315175, P.~R.~China}

\author{Bo Zheng}
\email{zhengbo@zju.edu.cn}
\affiliation{Department of Physics, Zhejiang University, Hangzhou 310027, P.~R.~China}
\affiliation{Collaborative Innovation Center of Advanced Microstructures, Nanjing 210093, P.~R.~China}

\begin{abstract}
With the help of transfer entropy, we analyze information flows between communities of complex networks. We show that the transfer entropy provides a coherent description of interactions between communities, including non-linear interactions. To put some flesh on the bare bones, we analyze transfer entropies between communities of five largest financial markets, represented as networks of interacting stocks. Additionally, we discuss information transfer of rare events, which is analyzed by R\'{e}nyi transfer entropy.

%it can acquire negative values, which can be interpreted as an additional uncertainty of marginal events transfer.
\end{abstract}

%\submitto{\NJP}
\maketitle

\section{Introduction}
Complex networks are the example of systems exhibiting a broad range of non-trivial emergent phenomena, including extreme shocks, memory effects or non-trivial correlation structure. These phenomena have been extensively studied by many sophisticated techniques:  power-laws~\cite{garbaix,yang,fabrizio}, auto- and cross-correlations~\cite{aste,longin,jiang,ramchand}, multi-scale models and multifractals~\cite{kaushik,schmitt,matteo,jizba1,ouyang}, complex dynamics~\cite{johnson,hilgers}, microscopic models~\cite{preis,chen,raberto} or graph theory~\cite{preis1,garlaschelli,squartini,mastrandrea}, just to name a few. One of the most important and most difficult tasks is to describe the collective behavior of interacting edges, and corresponding information flows between them.
%This is particularly important for detection of extreme shocks which can be transmitted to the rest of the network.

There exist several techniques for measurement of the information flows. To the most popular methods belong lagged cross-correlation~\cite{ramchand,chen} or Granger causality~\cite{granger,cheung}. Unfortunately, these quantities suffer from some shortcomings. The main disadvantage of cross-correlation is the fact that it is not possible to separate effects caused solely by the source time series and effects of the environment, e.g., effects of common information sources. For Granger causality, it is possible to detect information coming only from the source series. Nevertheless, both measures are based on linear models and can be therefore insensitive to nonlinear interactions.

These issues can be overcome by the introduction of information-based measures which can appropriately detect the information flows and identify its sources. Transfer entropy, introduced by Schreiber~\cite{schreiber}, is the model-free measure of information transfer between time series. It is based on famous Shannon information entropy~\cite{shannon} and has been successfully used in many applications~\cite{marschinski,kwon,palus,hlavackova,lungarella}. It has been shown that Granger causality and transfer entropy coincide in ``Gaussian world''~\cite{barnett}. Nevertheless, most complex networks are highly non-Gaussian and nonlinear. With the advent of generalized entropies~\cite{jizba2,tsallis,hanel}, many applications of entropy in thermodynamics, statistics and information theory found their natural generalizations. These generalizations can be utilized to describe complex and nonlinear dynamics more precisely. Following this scheme, Jizba \emph{et al.} introduced a new class of R\'{e}nyi transfer entropies~\cite{jizba3}. The specifics of R\'{e}nyi transfer entropy is given by the fact, that it is possible to focus on information transfer of certain parts of probability distributions. Since that, R\'{e}nyi transfer entropy has found several applications, e.g., in signal processing \cite{sarbu} or single-spin dynamics \cite{deng}. The topic of information transfer remains still a hot topic of ongoing research~\cite{choi,kang,teng}.

The main aim of this paper is to investigate methods for detection of information flows between communities in complex networks. As an example, we demonstrate the method on information flows  between communities in financial markets. Financial markets can be treated as complex networks with the inner structure of communities \cite{girvan,newman,rosvall,axelsson}. Typically, these communities correspond to business sectors \cite{jiang2}. So far, information flows have been measured only between financial markets~\cite{marschinski,kwon,jizba3} or between single stocks in one particular market. Our aim is not only to detect particular information flows but more generally to understand, how specific types of business sectors interact between each other. Such nonlinear interactions become even more prominent during some unusual trading period, as, e.g., financial crises, external driving, or company affairs.  Distribution of marginal events can be accentuated by choosing a proper $q$ for the R\'{e}nyi transfer entropy. All these aspects can help us understand different dynamics of particular markets.

The rest of the paper is organized as follows: Section \ref{sec: cc} discusses properties of the correlation function and its spectral decomposition into different modes. Section \ref{sec: cs} describes the algorithm for detection of the community structure in complex networks. Section \ref{sec: te} introduces the concept of transfer entropy and discusses its generalization based on R\'{e}nyi entropy. Section \ref{sec: fi} presents the structure of information transfer within and between communities in financial markets and discusses the differences between particular markets. Moreover, it compares flows of rare events by R\'{e}nyi transfer entropy. The last section is dedicated to conclusions.

\section{Correlation function and sector correlation}\label{sec: cc}
Correlation is one of the most important measures detecting the similarity of time series. For a pair of stationary time series $X(t)$, $Y(t)$, it is possible to introduce cross-correlation function as
\begin{equation}
C_{X,Y}(\tau) = \frac{\langle(X(t) - \mu_X)(Y(t-\tau)-\mu_Y)\rangle}{\sigma_X \sigma_Y},
\end{equation}
where $\mu_X$ and $\mu_Y$ are mean values of each time series and $\sigma_X$, $\sigma_Y$ are their standard deviations. Naturally, $C \in [-1,1]$. We can distinguish three different cases. First, for $X = Y$ we obtain auto-correlation function. It is used to detect dependencies in the time series. The characteristic correlation time is for financial returns approximately 5 min in most of the financial markets,as shown, e.g., in~\cite{stanley}.

Second, for $\tau=0$ we get equal time cross-correlation, which can be treated as a measure of similarity between two series. It cannot be used as the measure of information flow because it lacks directionality and can be affected by effects of the environment. Nevertheless, it is used in many standard techniques, including detection of community structure, as shown in section \ref{sec: cs}.

Finally, for $X \neq Y$ and $\tau > 0$, we get lagged cross-correlation. It has clear directionality meaning. Unfortunately, it is hard to distinguish between causality and other forms of dependence, and it may not be sensitive to nonlinear interactions. In the case of a noisy system like financial markets, cross-correlations decay in terms of minutes, and it is not possible to detect the interactions beyond this scale in most cases. These issues serve as a motivation for the introduction of measures based on information theory, as, e.g., mutual information or transfer entropy. We present the main results of information theory in section \ref{sec: te}.

\subsection{Mode decomposition of correlation matrix}
In a noisy environment with external information sources, correlation matrix contains not only the information about interactions between the time series, but also global market movement and noise fluctuations. Let us define the correlation matrix $\mathbb{C}$ between time series $X_i(t)$ and $X_j(t)$ as
\begin{equation}
\mathbb{C}_{ij} = C_{X_i,X_j}(\tau=0)\, .
\end{equation}
The spectrum of the correlation matrix $\mathbb{C}$ is real because of the symmetry. The matrix can be represented via its spectral decomposition
\begin{equation}
\mathbb{C} = \sum_\alpha \lambda_\alpha \mathbf{u}_\alpha \otimes \mathbf{u}_\alpha ,
\end{equation}
where $\lambda_\alpha$ are the eigenvalues, and $\mathbf{u}_\alpha$ are the eigenvectors. The correlation matrix for $N$ uncorrelated time series in finite time $T$ is known as Wishart matrix and the distribution of its eigenvalues is for $N \rightarrow \infty$, $T \rightarrow \infty$, $Q = T/N \gg 1$ given by \cite{sengupta,dyson}
\begin{equation}
P(\lambda) = \frac{Q}{2\pi} \frac{\sqrt{(\lambda_{max}-\lambda)(\lambda-\lambda_{min}{})}}{\lambda} ,
\end{equation}
where $\lambda_{max/min} = \left[1 \pm (1/Q)^{1/2}\right]^2$. Thus, the eigenvalues, for which holds $\lambda > \lambda_{max}$, represent the non-random interactions of the system. Let us order the eigenvalues, so $\lambda_i \geq \lambda_{i+1}$. In many systems, like financial markets, the largest eigenvalue represents the overall system (market) mode \cite{jiang2}. Local interactions can be described by the sector correlation $\mathbb{C}^{sec}$ which is defined as
\begin{equation}
\mathbb{C}^{sec} = \sum_{\alpha=2}^{\alpha_{max}} \lambda_\alpha \mathbf{u}_\alpha \otimes \mathbf{u}_\alpha ,
\end{equation}
where $\lambda_{\alpha_{max}}$ is the smallest eigenvalue larger than $\lambda_{max}$.

\section{Community structure in complex networks}\label{sec: cs}
Correlation matrix (or sector correlation matrix) contains the full information about all interactions of the network constituted by particular stocks. Unfortunately, the number of links is $2^{{N}\choose{2}}$, which becomes a huge number for large networks, while only a small fraction of links play a relevant role in the dynamics of the network. Therefore, it is not only desirable but also necessary to discard most of the links and keep the most significant ones. For this end, several simple algorithms have been proposed. Minimal spanning tree (MST) algorithm \cite{kruskal,prim} is based on a simple idea. Having correlations ordered in the descending order, we add the link to the filtered graph. If the addition of the edge would create the circle, we do not add it and move to the next edge, until we obtain the fully connected graph. Such a graph contains $N$ nodes and $N-1$ edges and maximizes the aggregated correlation of the graph.

This simple algorithm has nevertheless a distinctive flaw. Because of the tree structure, significant correlations which would create a circle are omitted. Such circles can describe a small set of stocks strongly interacting with each other. Therefore, a generalization of MST method was proposed in Ref.~\cite{tumminello}. The method is called Planar maximally filtered graph (PMFG). It is based on a similar idea as MST algorithm, but the condition of no graph circles is replaced by planarity condition, i.e., that it is possible to embed the graph in a plane. The generated graph can be viewed as a triangulation of the plane. It is possible to show, that the PMFG graph consists of $3N-6$ edges.

Complex networks usually consist of nontrivial structures of highly connected nodes, which create communities. Interactions in the communities are typically very strong, while interactions between communities vary from relatively strong to very weak. The importance of community detection was pointed out by many authors \cite{girvan,newman}. One of the successful methods, which is used for community detection is called InfoMap. It has been introduced in Refs. \cite{rosvall,axelsson} and is based on the optimal compression of information flows in the networks. The main idea is to minimize the average code describing the random walk of the network. Typically, the walker remains in one community for a long time and then suddenly jumps to another community. In financial markets, the approach was successfully used, together with correlation decomposition method, in Ref. \cite{jiang2} to reveal the structure of business sectors.

\section{Transfer entropy}\label{sec: te}

The concept of entropy was introduced by Shannon~\cite{shannon} in 1948. According to Campbell coding theorem~\cite{cambell}, it represents the minimal amount of binary information necessary to encode a message. It can be expressed as
\begin{equation}
H(X) = - \sum_x p(x) \log_2 p(x),
\end{equation}
where $p(x)$ is the probability distribution of occurrence of symbols from the alphabet $\{A_i\}_{i=1}^S$. We denote the discrete random variable as $X$. Let us remind that the information entropy is closely related to well-known thermodynamic Boltzmann-Gibbs entropy via the multiplicative factor $k_B$ (Boltzmann constant). It is important to say that the term $H(X)$ represents only information contained in $X$. Analogously, information contained in $X \cup Y$ can be represented by the joint entropy $H(X \cup Y)$ given by the joint distribution $p(x,y)$. If $X$ is statistically independent of $Y$, the joint entropy is just $H(X \cup Y) = H(X) + H(Y)$. In agreement with Khinchin axiomatic definition of Shannon entropy~\cite{khinchin}, conditional entropy can be introduced as

%\numparts
\begin{align}\label{eqconditional}
H(Y|X) = H(X \cup Y) - H(X).
 %      &=& \sum_{x,y} p(x,y) \log_2 p(y|x)
\end{align}
%\endnumparts
%where $p(y|x) = p(x,y)/p(x)$ is the conditional distribution.
This allows us to introduce mutual information of $X$ and $Y$ as
%\numparts
\begin{align}\label{eqmutual}
I(X;Y) = H(X) - H(X|Y) = H(Y) - H(Y|X)\nonumber\\
 =H(X) + H(Y) - H(X \cup Y).
       %&=& \sum_{x,y} p(x,y) \log_2 \frac{p(x,y)}{p(x)p(y)}\, .
\end{align}
%\endnumparts
%$p(y)$ is the distribution belonging to $Y$.
%We can straightforwardly generalize the mutual information to more than two variables following \cref{eqmutual}.
%% lagged mutual information
Information flow from time series $Y(t)$ to time series $X(t)$ can be introduced as lagged mutual information $I(X(t);Y(t-\tau))$. Nevertheless, although directional, it also includes mutual information, which is induced by a common external source, e.g., when both $X$ and $Y$ are statistically dependent on a common random variable $Z$. In this case, it is convenient to express the mutual information conditioned by the source $Z$ by
\begin{equation}\label{eqmutualconditional}
I(X;Y|Z) = H(X|Z) - H(X|Y \cup Z) .
\end{equation}
Let us consider discrete time series $X(t)$, $Y(t)$ with a constant time lag $\tau$. Let us denote $x_{m+1} = X(m+1)$, $\mathcal{X}_m = \{X(m),\dots,X(1)\}$ and $\mathcal{Y}_m^l = \{Y(m),\dots,Y(m-l+1)\}$. For stationary series is $p(\mathcal{X}_m) \sim p(\mathcal{X}_{m+t}^m)$, and similarly $p(\mathcal{Y}_m^l) \sim p(\mathcal{Y}_{m+t}^{l})$. Shannon transfer entropy (STE) is defined as the conditional mutual information
\begin{equation}
T_{Y \rightarrow X}(m,l) = I(x_{m+1};\mathcal{Y}_m^l|\mathcal{X}_m)\, .
\end{equation}
The definition is clearly directional. What more, it takes into account only the dependency whose origin is in source series $Y$. Dependencies caused by common external sources are not taken into account. It is possible to write it explicitly as
\begin{eqnarray}
T_{Y \rightarrow X}(m,l)=-\sum p(x_{m+1},\mathcal{X}_m,\mathcal{Y}_m^l) \log_2 \frac{p(x_{m+1}|\mathcal{X}_m)}{p(x_{m+1}|\mathcal{X}_m,\mathcal{Y}_m^l)}\, .
\end{eqnarray}

Naturally, transfer entropy depends on history indices $m$ and $l$. For Markov processes of order $m$ and $l$ is sufficient to use the history up to the order of the processes. Unfortunately, for non-Markovian processes, i.e., processes with the long history is this not possible. Ideally, one should take into account the whole history of the series, i.e., $m \rightarrow \infty$ to find a stable value independent of $m$. Unfortunately, this is not possible due to the finite size of the dataset. It might be an issue even for very long datasets because the number of possible states grows as $S^{m+l}$, which also affects computational time and accuracy. Therefore, Marchinski and Kantz introduced the effective transfer entropy~\cite{marschinski} to avoid the problems with finite-size effects as
\begin{equation}
T^{eff}_{Y \rightarrow X}(m,l) = T_{Y \rightarrow X}(m,l) - T_{Y_{\mathrm{shuffled}} \rightarrow X}(m,l)\, .
\end{equation}
Typically, it is desirable to choose $m$ as large as possible and $l=1$~\cite{marschinski}. For the daily data, it is convenient to choose also $m=1$, because the typical scales influencing the dynamics of returns are usually below the scale of days. Moreover, the datasets typically contain data over several years or maximally decades and larger $m$ would require much more data in order to suppress the finite-size effects.

\subsection{R\'{e}nyi transfer entropy}\label{sec: re}
Shannon entropy is not the only one functional fulfilling the Khichin additivity axiom for conditional entropy. It has been shown by R\'{e}nyi~\cite{renyi} that there exists the whole class of entropies, characterized by parameter $q > 0$, which can be expressed as
\begin{equation}
S_q(X) = \frac{1}{1-q}\log_2 \sum_x p(x)^q\, .
\end{equation}
In the limit $q \rightarrow 1$, we recover the Shannon entropy. One can formulate (similarly to Shannon entropy) the coding theorem for R\'{e}nyi entropy. It describes the minimal cost necessary to encode a message when the cost is an exponential function of the message length~\cite{bercher}. R\'{e}nyi entropy is closely related to multifractal systems~\cite{jizba2,hentschel} and escort distributions, which can be defined as
\begin{equation}
\rho_q(x) = \frac{p(x)^q}{\sum_{x'} p(x')^q}\, .
\end{equation}
This class of distributions conforming a group was originally introduced in connection with chaotic systems~\cite{beck}. They are also called ``zooming distributions'', because for $q<1$, they highlight tail parts of the distribution, while for $q>1$, they suppress them and accentuate central parts of the distribution.

Contrary to other generalized entropies (for details, see, e.g.,~\cite{tsallis,hanel}), the relevant operational definitions of mutual information (\cref{eqmutual}) and conditional mutual information (\cref{eqmutualconditional}) remain R\'{e}nyi entropy valid, which is the consequence of additivity. Consequently, R\'{e}nyi transfer entropy (RTE) can be expressed as
\begin{align}
T_{q;Y \rightarrow X}(m,l)=  \frac{1}{1-q} \log_2 \frac{\sum \rho_q(\mathcal{X}_m) \, p^q(x_{m+1}|\mathcal{X}_m)}{\sum \rho_q (x_{m+1}, \mathcal{X}_m,\mathcal{Y}^l_m) \, p^q(x_{m+1}|\mathcal{X}_m,\mathcal{Y}^l_m)}
\end{align}
which for $q \rightarrow 1$ again boils down to STE.

The main difference between STE and RTE is that we can focus on different parts of distributions by varying the parameter $q$. Contrary to STE, RTE can be negative, which is equivalent to the situation when
\begin{equation}
S_q(x_{m+1}|\mathcal{X}_m \cup \mathcal{Y}_m^l) \geq S_q(x_{m+1}|\mathcal{X}_m)\, .
\end{equation}
This paradoxical behavior brings a desirable information that the knowledge of historical values $X$ and $Y$ reveals an extra risk to certain parts of distribution coming from nonlinear interaction between stocks. Consequently, RTE cannot be interpreted as a strength of information. On the other hand, it is possible to understand the negative RTE as the presence of emergent collective interactions among the stocks, leading to increased complexity of the network.

We are typically interested in information transfer of the tail parts of the distribution, i.e., swan-like events. Naturally, these events do not occur so often, but they can remarkably affect the whole network. These events can spread in the network and cause the avalanche effect. For detection of swan-like events, it is possible to adjust the parameter $q$ to values smaller than one. Because for small values is the method very sensitive to errors, one usually chooses a compromise value, e.g., $q=0.75$.

\section{Information transfer between business sectors in financial markets}\label{sec: fi}
\begin{table}[t]
\begin{tabular}{|l||c|c|c|c|c|}
  \hline
  market & New York & London& Tokyo & Shanghai & Hong Kong\\
  index & S\&P 500 & FTSE AS & NKY 225 & SSE 300 & HSI Comp. \\
  \hline
  stocks & 485 & 527 & 185 & 283 & 411 \\
  av. length (days) & 3905  & 2206 & 2205 & 2929 & 3677\\
  $\overline{C}(0)$ & 0.2122 & 0.0390 & 0.2185 & 0.2601 & 0.1448\\
   $\overline{C}(1)$ & -0.0093 & 0.0069 & 0.0002 & 0.0193 & 0.0229 \\
  $\overline{STE}$ & 0.0047  & 0.0008 & 0.0027 & 0.0030 & 0.0019 \\
  $\overline{RTE}_{0.75}$ & -0.0260  & -0.0062 & 0.0027 & -0.0218 & -0.0162\\
  communities & 22 & 22 & 10 & 16 & 25 \\
  \hline
\end{tabular}
\caption{Statistical properties of investigated markets. For each market, the table contains number of investigated stocks, average length of trading period, equal-time cross-correlation, lagged cross-correlation, Shannon and R\'{e}nyi transfer entropy and number of communities obtained from InfoMap algorithm.}
\label{tab: fm}
\end{table}
In order to give an example of a system with complex interactions, let us turn our attention to the analysis of information transfer between business sectors in financial markets. We investigate five largest stock exchanges (SE) according to the market capitalization -- New York SE, London SE, Tokyo SE, Shanghai SE and Hong Kong SE. These markets have sufficiently rich structure and contain stocks from various business sectors. Each market is represented by a set of assets included in one of the main market indices. For each market, we investigate the period of the last 10-16 years (ending in 2016). We include only stocks that have been traded at least 1000 days (approx. 4 years). Basic statistics of all markets, including the number of investigated stocks and average length the time series, is contained in Table~\ref{tab: fm}. Each stock is represented by its price $S_i(t)$. Daily returns are given as
\begin{equation}
r_i(t) = \ln S_i(t) - \ln S_i(t-1).
\end{equation}
For each market, we calculate the correlation matrix of daily returns and extract the sector mode correlation from the spectral decomposition. The sector mode correlation can be used for the definition of the adjacency matrix of the financial market network, which is usually defined as $\mathbb{A}_{ij}= 1-\mathbb{C}_{ij}$. %The adjacency matrix contains the full information about the network.
We filter the edges with the help of PMFG method and keep only edges with the lowest distance (largest correlation).

Using the InfoMap algorithm, we determine the communities of each market. These communities correspond very well to business sectors. Nevertheless, for some markets, it is possible to find communities consisting of two anti-correlated subsectors~\cite{jiang2}. This evidence is confirmed in all investigated markets. The only exception is provided by large conglomerates operating in several business sectors. In some cases, like for Hong Kong SE, the business sectors are also influenced by the country of residence. The main sectors consist of eight types of companies - basic materials, consumer goods, financial services, industry, services, technology, healthcare,and utilities. The communities often correspond to a specific subsector/industry, list of all industries and their abbreviations can be found in \ref{sec: ap}. Additionally, the method identifies flows between communities according to their mutual correlations.

For each market, we calculate average transfer entropy between communities. In all calculations, we use the 3-letter alphabet with division into intervals $(5\%,90\%,5\%)$, i.e., the central part and significant falls and rises. This division enables us to focus on the large market movements, and filter out the noisy background of the system. Similarly to correlation network, we obtain the full directed network of information flows between communities. Again, only a fraction of these flows is significant. For this end, we use the bootstrap methods \cite{dimpfl} and an appropriate threshold.
% which give average transfer entropies between shuffled time series (similarly to effective transfer entropy). The appropriate threshold is chosen as 2 standard deviations above the mean TE.

Before analyzing each market separately, let us note several general remarks.  In all cases, the correlation network is remarkably different from the information flow network for all markets. While community structure corresponds to the business structure determined by frequent interactions on the market, information flows reveal the structure hidden under regular interactions. Generally, information flows are strongest in two situations: first, between financial communities (e.g., banks, investment institutions or real estates) and second, large enterprises belonging to the key industry sectors of the particular country (car manufacturers in Germany or steel production industry in China). To be more specific, let us turn our attention to individual markets.

\begin{figure}[t]
\includegraphics[width=16cm]{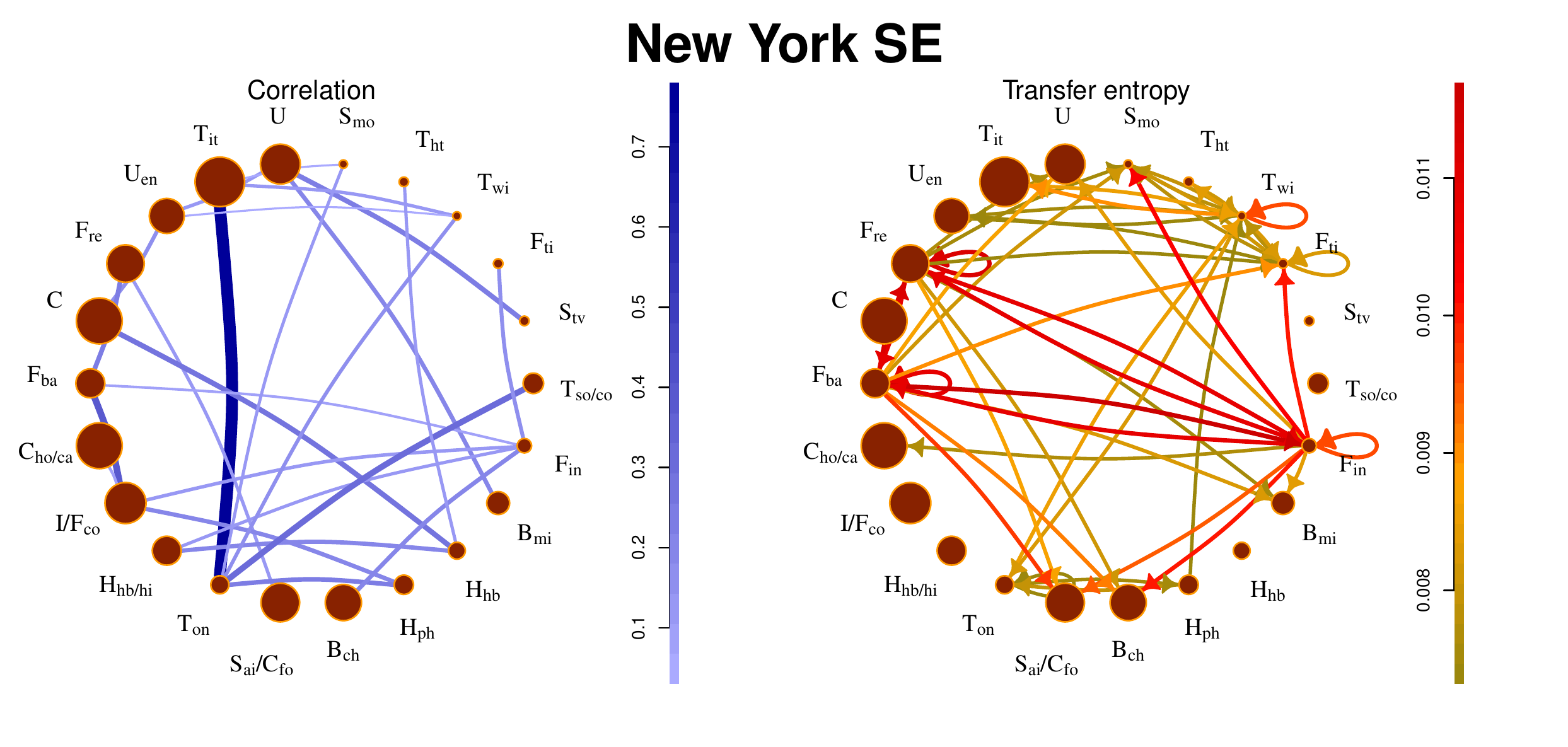}
\caption{Comparison of correlation network and transfer entropy flows between communities of New York SE. The strongest correlations are observed for information technology sectors and consumer goods sectors, for financial sectors are correlations slightly weaker. On the other hand, information flows between financial sectors are the most significant, influencing also several other sectors e.g., movie production sector. Interestingly, information flows between individual stocks within the community, represented by the closed loops, are also significant for the financial sectors. }
\label{fig1}
\end{figure}

\paragraph{New York SE:}
this market exhibits the largest information flows among all investigated markets. We observe the strongest flows between financial communities. They also affect the other communities (movie production, consumption, etc.). It is not a surprising fact because real estate sector is in the U.S. tightly connected with banking and investment sector and the financial crisis 2008 started in real estate sector and spread to the banking sector. These connections are also present in the correlation picture, but the strongest correlations are observed in the technology industry. The comparison between correlation network and transfer entropy flows is depicted in \Cref{fig1}. The stocks traded at New York SE exhibit the strongest correlations, largest information flows, and most significant complexity (as discussed in section \ref{sec: it}).

\begin{figure}[t]
\includegraphics[width=8cm]{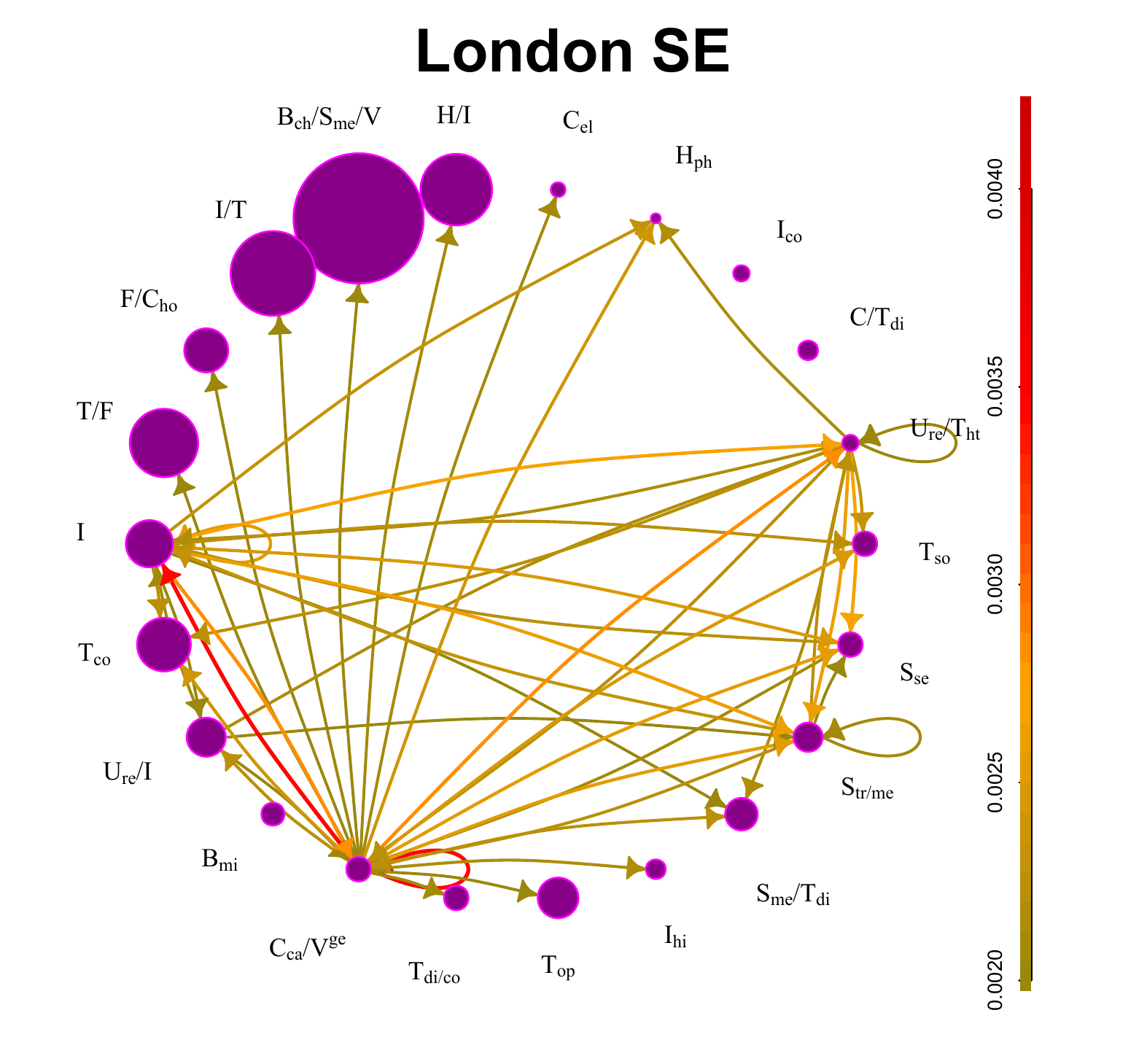}
\includegraphics[width=8cm]{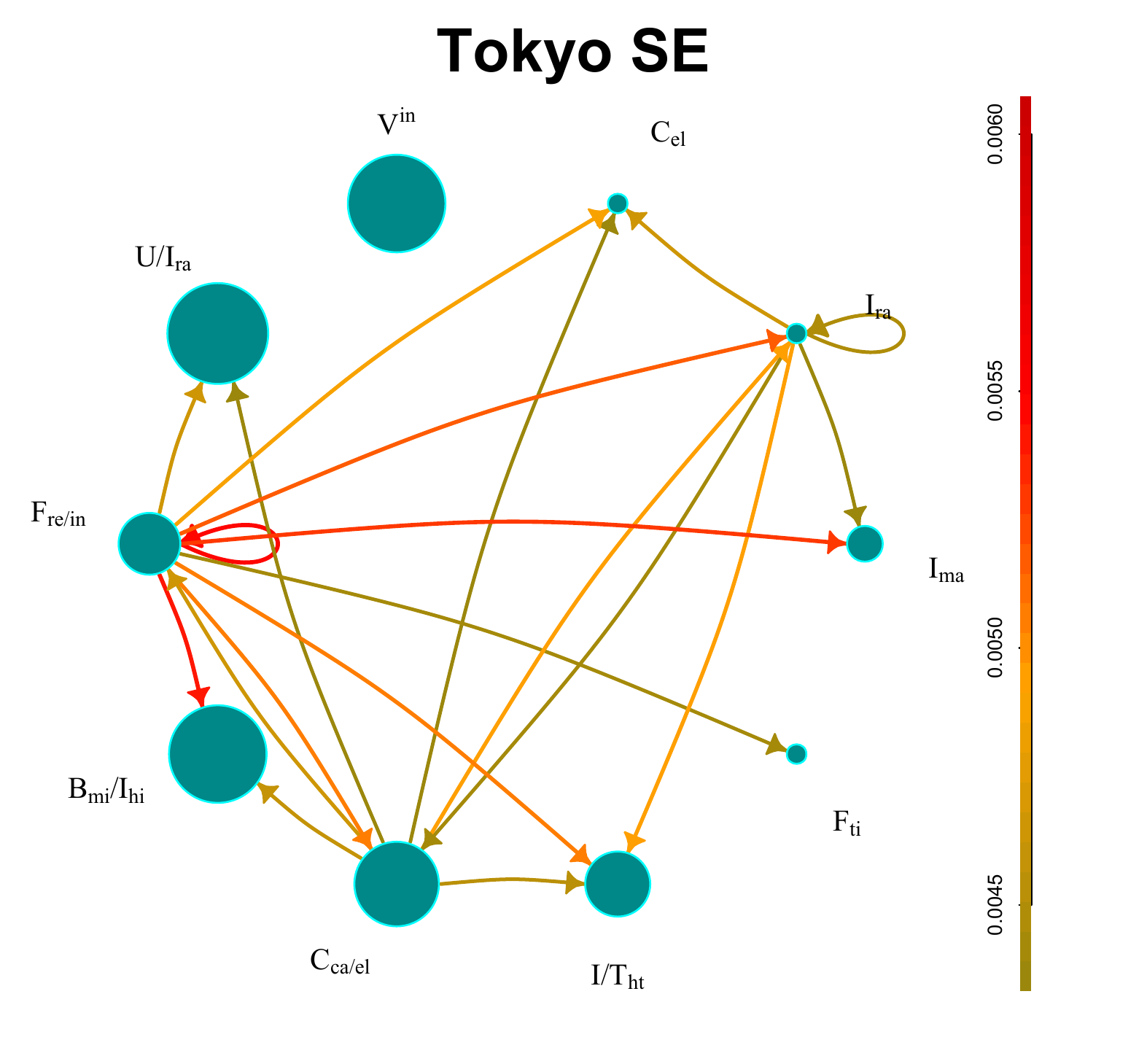}
\includegraphics[width=8cm]{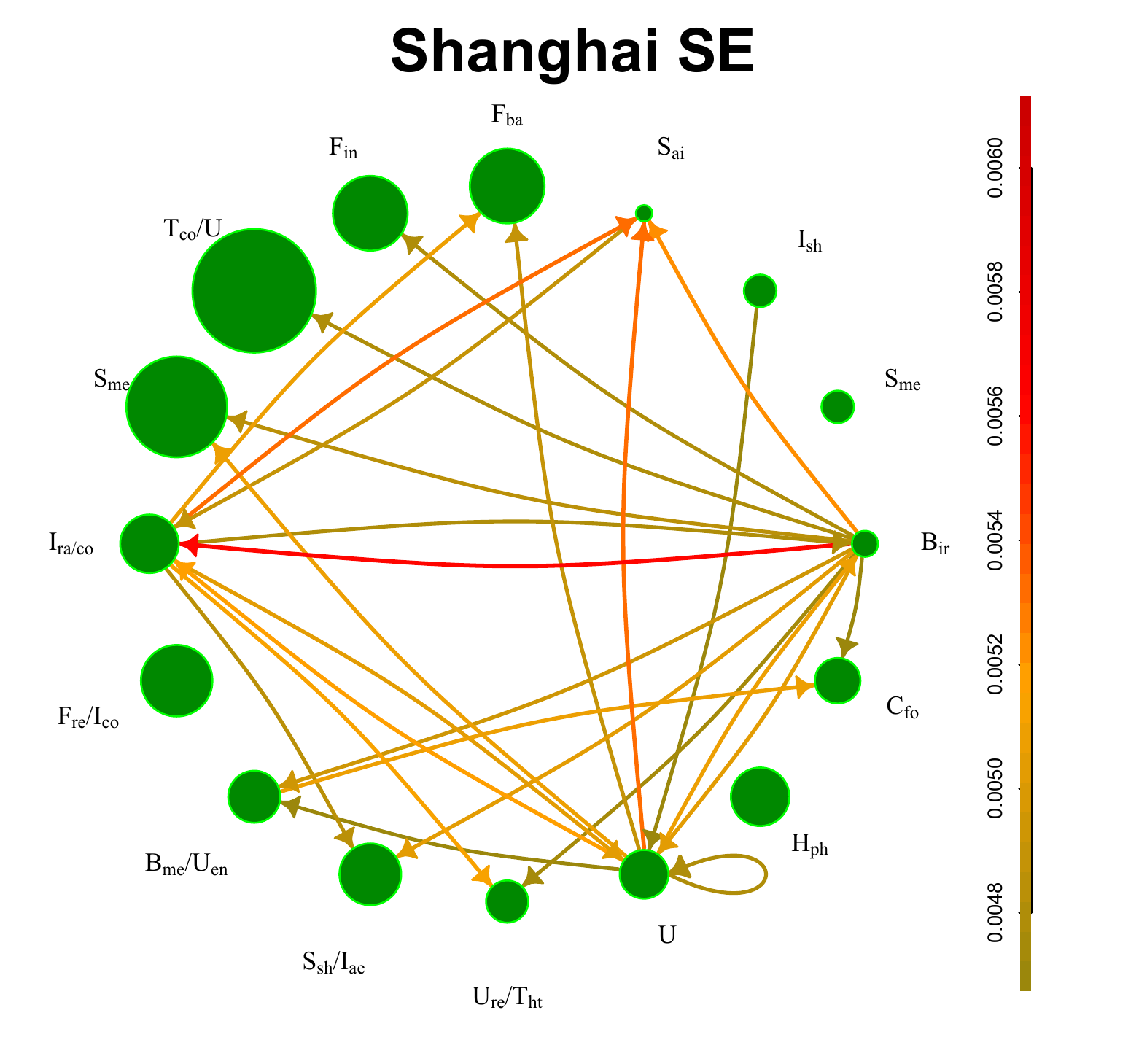}
\includegraphics[width=8cm]{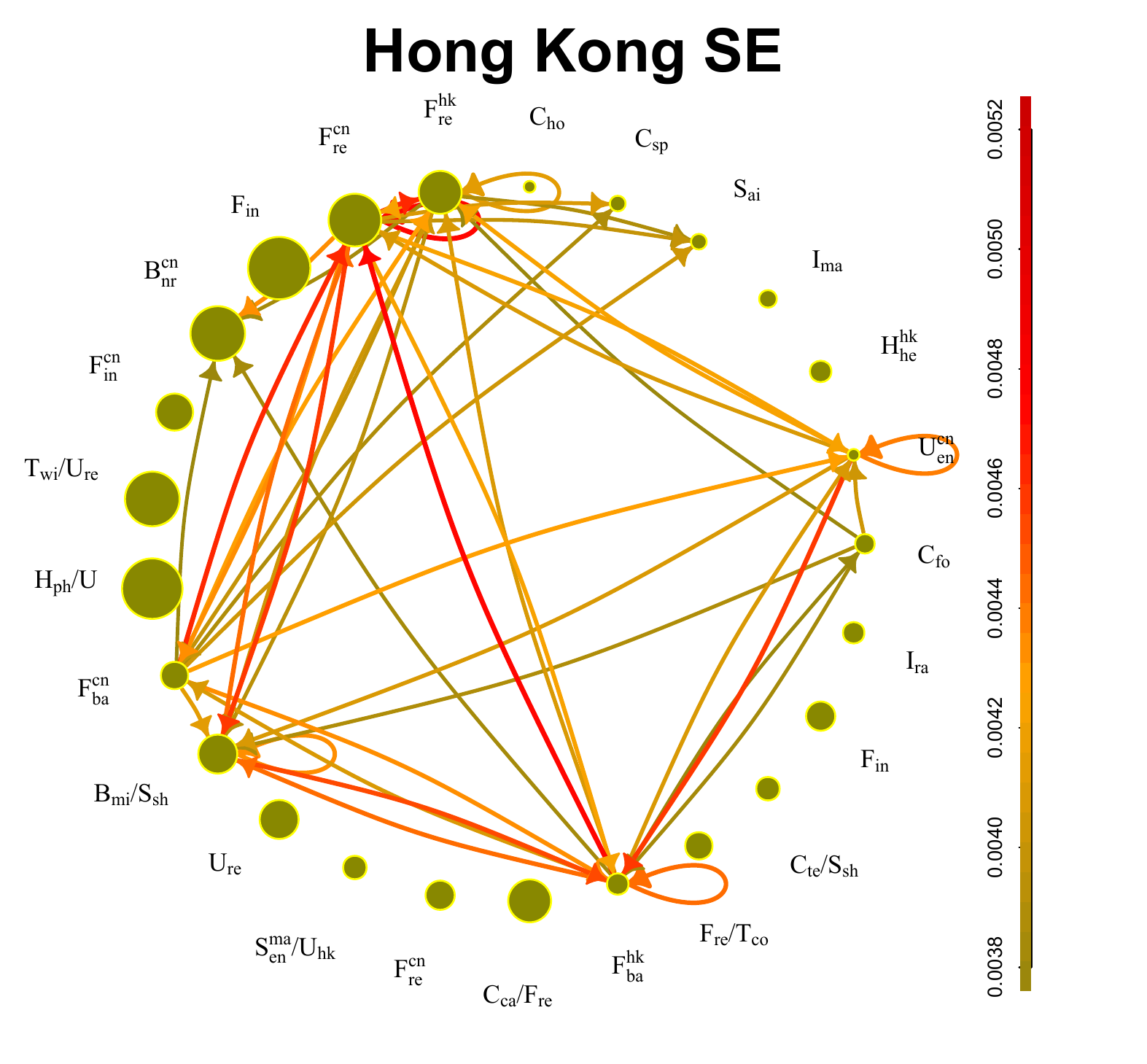}
\caption{Information flows between communities for London SE, Tokyo SE, Shanghai SE and Hong Kong SE. The strongest flows can be observed between financial sectors and between key industry sectors (German car manufacturers at London SE or steel production sector at Shanghai SE). Typically, the key industry sectors play the major role in the economy of the country. In some cases, as, e.g., Hong Kong SE, the community structure is not influenced by the business sector but also by the country of origin.}
\label{fig2}
\end{figure}

\paragraph{London SE:}
contrary to New York SE, interactions at London SE are much weaker. Stocks are very weakly correlated, and there are not many large information flows (see \cref{tab: fm}). The reason may be found in the structure of the market: most companies are industrial and technology companies producing various types of high-tech products or large multi-sector conglomerates. Their performance is not much influenced by the other companies in the market. The only community with significant outflows contains major German financial and service companies (Deutsche Bank, Deutsche Post, Allianz, Deutsche B\"{o}rse, Lufthansa) and large German car manufacturers (BMW, Daimler Chrysler, Volkswagen). The information flows of London SE and all remaining markets are depicted in \Cref{fig2}.

\paragraph{Tokyo SE:}
the market has relatively small number of large communities. Strongest flows can be found between financial sectors. There are also industry sectors with significant flows, as, e.g., electronics productions, or railway construction. Interestingly, the sector containing all international companies remains isolated.

\paragraph{Shanghai SE:}
there can be identified two sources of information. Particularly, the railway construction industry and steel production. China the largest steel producer and exporter and the sector has a great impact on the other areas of Chinese industry. On the other hand, financial companies do not produce strongest flows, also because the financial companies are also listed in Hong Kong SE.

\paragraph{Hong Kong SE:}
Interactions on the Hong Kong SE are influenced not only by affiliation to business sectors but also by the country of origin. Companies from mainland China and Hong Kong are contained in approximately same amount. Since Hong Kong market includes the large financial sector from both mainland China and Hong Kong, it is not surprising that strongest flows are among financial sectors. The flows are strong also among sectors with different country of origin.

\subsection{Information transfer of rare events and market complexity}\label{sec: it}
\begin{figure}[t]
\includegraphics[width=16cm]{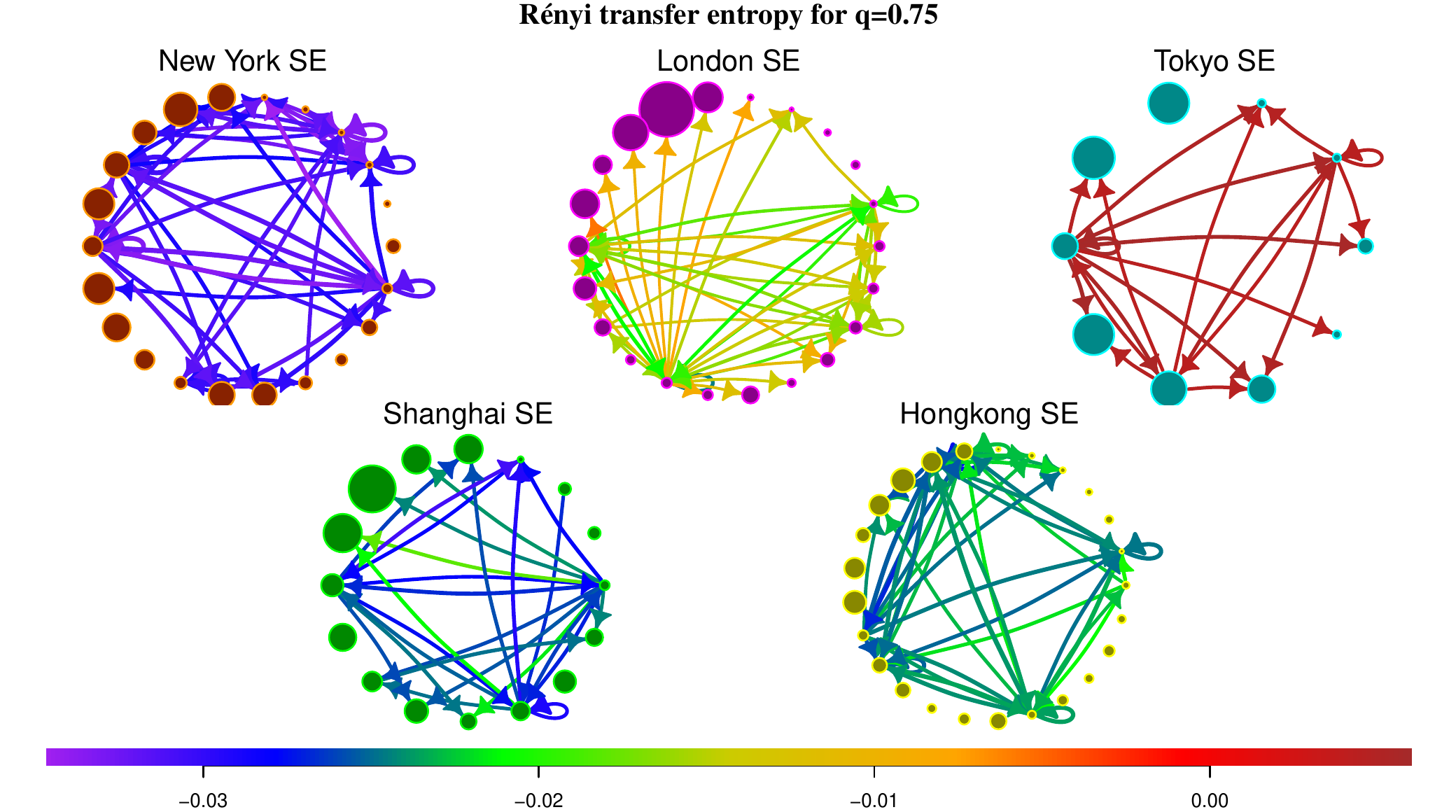}
\caption{R\'{e}nyi transfer entropy in five financial markets. We compare the most significant information flows obtained by the Shannon transfer entropies and color them by corresponding R\'{e}nyi transfer entropies. Since for $q=0.75$ describes RTE the information transfer of rare events, we obtain a different view. Negative RTE corresponds to the fact that knowledge historical values of source series reveals an extra risk for the tail parts of recipient, which makes the interaction more complex.}
\label{fig3}
\end{figure}
Let us focus on RTE and transfer of rare events. As discussed in \Cref{sec: re}, RTE for $q < 1$ accentuates transfer of  marginal events. We calculate the average RTE between communities for all markets and focus on most significant flows, i.e., flows with largest STE (see \Cref{fig3}). In Ref.~\cite{jizba3} have been analyzed R\'{e}nyi information flows between indices of different financial markets. In most cases is RTE positive. The only exception is the information transfer between indices S\&P 500, DJIA and NYSE 100. All these indices are created from stocks of New York SE, which points to the fact that interactions in New York SE are very sensitive to marginal events, which is also confirmed by our analysis. New York SE exhibits the lowest values of RTE for $q=0.75$ among all markets (average values of STE and RTE are listed in \Cref{tab: fm}). It reflects the fact that New York SE is a well-developed market with the complex structure which recently passed through the large financial crisis (the crisis and post-crisis data constitute the major part of the investigated period). On the contrary, London SE and especially Tokyo SE have much higher values of RTE, for some flows in London SE and all flows in Tokyo SE is the RTE even positive. Shanghai SE and Hong Kong SE are somewhere between these two types of behavior. Generally, the information transfer of swan-like events between markets is much more predictable than within financial markets, especially for New York SE.

\section{Conclusions and perspectives}
%Description of information flows in complex networks is an important step towards understanding the structure of interactions in the system. In many real systems, as e.g., in financial markets, the interactions are highly non-linear. Adequate description should take into account the complexity of interactions. Usual techniques based on linear models can be insufficient. Shannon transfer entropy, on the other hand, is a model-free measure based on concepts of information theory which describes the strength of information flows, including also non-linear interactions.

The main aim of this paper was to investigate information flows between communities of complex networks. Information flows can be measured by transfer entropy, a model-free measure quantifying the amount of information transmitted from the source time series to the target time series. It can successfully describe complex systems with nonlinear interactions. As an example, we have analyzed five largest financial markets. We find that the strongest flows are observed for financial sectors and key industry sectors (e.g., German car manufacturers at London SE or steel producers at Shanghai SE). On the other hand, sectors with high correlations, as technology sectors or consumer goods, exhibit much weaker information flows. It is caused by the fact that the former sectors produce significant information transfer of marginal events, which becomes much more important in the transfer entropy picture.

To emphasize the importance of rare events transfer, we introduced the  R\'{e}nyi transfer entropy which enables one to study information flows between specific parts of probability distributions. R\'{e}nyi transfer entropy can acquire negative values, which can be interpreted as an additional risk (or uncertainty) for specific parts of distribution of the target series. Negative R\'{e}nyi transfer entropy can be interpreted as the increased complexity of the network. We have compared R\'{e}nyi transfer entropy for $q=0.75$ among the five example markets. As a result, some markets, especially New York SE, exhibit negative R\'{e}nyi transfer entropy for most flows, which signalizes that the transfer of rare events transfer is nonlinear and unpredictable - the network is complex. It should be taken into account when designing models of risk spread and in the modeling of swan-like events.

Dynamics of information flows measured by transfer entropy provides a different description of complex financial networks, when compared with interactions measured by correlations. Therefore, complex networks based solely on information transfer would provide a novel approach to understanding of complex networks dynamics. Because information flows are directional, it will be necessary to adjust the procedures to be able to deal with directed graphs. Investigation of communities based on directed transfer entropy-based networks is the subject of ongoing research.

\section*{Acknowledgements}
Authors want to thank Petr Jizba and Martin Prok\v{s} for helpful conversations. B. Z. and X. F. J. acknowledge the financial support from NNSF of China, grant No. 11505099 and No. 11375149, and the financial support from Zhejiang Provincial Non-Profit Fund for Applied Research, grant No.
No. 2016C33248. J. K. acknowledges the financial support from the Czech Science Foundation, grant No. 17-33812L.

\appendix

\section{Sectors and Industries}\label{sec: ap}
\begin{scriptsize}
\begin{tabular}{ | l | l | l | l | l | l | }
\hline
	\textbf{B} & \textbf{Basic materials} & \textbf{C} & \textbf{Consumer goods} & \textbf{F} & \textbf{Financial services} \\ \hline
	$B_{ch}$ & chemistry & $C_{ca}$ & car manufacture & $F_{ba}$ & banks \\
	$B_{ir}$ & iron \& steel & $C_{el}$ & Electronics & $F_{co}$ & consulting services \\
	$B_{me}$ & metals & $C_{fo}$ & food & $F_{hi}$ & health insurance \\
	$B_{mi}$ & mining & $C_{sp}$ & sport \& lifestyle & $F_{in}$ & investment services \\
	$B_{nr}$ & natural resources & $C_{te}$ & textile & $F_{re}$ & real estate \\
	 &  & $C_{ho}$ & household & $F_{ti}$ & travel \& accident insurance \\ \hline
	\textbf{I} &\textbf{ Industrial goods} & \textbf{S} & \textbf{Services} & \textbf{T} & \textbf{Technology} \\ \hline
	$I_{ae}$ & aerospace \& defense & $S_{ai}$ & airlines & $T_{co}$ & communications \\
	$I_{ag}$ & agriculture industry & $S_{en}$ & entertainment & $T_{di}$ & digital services \\
	$I_{hi}$ & heavy industry & $S_{me}$ & media & $T_{ht}$ & high-tech industry \\
	$I_{in}$ & infrastructure & $S_{mo}$ & movie production & $T_{it}$ & information technology \\
	$I_{ma}$ & machinery & $S_{se}$ & security services & $T_{on}$ & online services \\
	$I_{ra}$ & railway construction & $S_{sh}$ & shipping & $T_{op}$ & optical \& nano technology \\
	$I_{sh}$ & ship construction & $S_{tr}$ & transportation & $T_{so}$ & software \\
	$_{Ive}$ & vehicle industry & $S_{tv}$ & television & $T_{wi}$ & wireless services \\ \hline
	\textbf{H} & \textbf{Healthcare} & \textbf{U} & \textbf{Utilities} &  & \textbf{Country }\\ \hline
	$H_{hb}$ & health \& beauty & $U_{en}$ & energy & ${}^{cn}$ & China \\
	$H_{me}$ & medical equipment & $U_{ga}$ & gas \& oil & ${}^{hk}$ & Hongkong \\
	$H_{ph}$ & pharmacy & $U_{re}$ & renewable energy & ${}^{ma}$ & Macau \\
	 &  &  &  & ${}^{in}$ & international \\
	 &  & \textbf{V} & \textbf{Various and conglomerates} & ${}^{ge}$ & major German companies \\ \hline
\end{tabular}
\end{scriptsize}

\section*{References}
\bibliographystyle{unsrt}
\bibliography{references}

\end{document}